\documentclass{emulateapj}
\usepackage{apjfonts}

\newcommand{\Fig}[1]{Figure~\ref{#1}}

\newcommand{\Msun}{\ensuremath{{\rm M}_\sun}}
\newcommand{\Zsun}{\ensuremath{{\rm Z}_\sun}}
\newcommand{\pow}[2]{\ensuremath{#1 \times 10^{#2}}}
\newcommand{\err}[3]{\ensuremath{#1_{-#2}^{+#3}}}

\shorttitle{Ly$\alpha$ Emitting Galaxies at $z = 3.1$}
\shortauthors{Lai et al.}

\begin{document}

\newcommand{\zee}{z}

\title{Spitzer Constraints on the Stellar Populations of Lyman-alpha
  Emitting Galaxies at $\zee = 3.1$}

\author{Kamson Lai\altaffilmark{1}, Jia-Sheng Huang\altaffilmark{1},
  Giovanni Fazio\altaffilmark{1}, Eric Gawiser\altaffilmark{2,3}, 
  Robin Ciardullo\altaffilmark{4}, Maaike Damen\altaffilmark{5},
  Marijn Franx\altaffilmark{5}, Caryl Gronwall\altaffilmark{4},
  Ivo Labbe\altaffilmark{6}, Georgios Magdis\altaffilmark{7},
  Pieter van Dokkum\altaffilmark{2}
}

\altaffiltext{1}{Harvard-Smithsonian Center for Astrophysics, 60
  Garden Street, Cambridge, MA 02138, USA; klai@cfa.harvard.edu}

\altaffiltext{2}{Yale Astronomy Department and Yale Center for Astronomy and 
  Astrophysics, Yale University, P.O. Box 208121, New Haven, CT 06520,
  USA}

\altaffiltext{3}{Department of Physics and Astronomy, Rutgers
  University, 136 Frelinghuysen Road, Piscataway, NJ 08854-8019}

\altaffiltext{4}{Department of Astronomy \& Astrophysics, The
  Pennsylvania State University, Davey Lab, University Park, PA 16802,
  USA}

\altaffiltext{5}{Sterrewacht Leiden, Leiden University, Postbus 9513, 
  NL-2300 RA Leiden, Netherlands}

\altaffiltext{6}{Carnegie Observatories, 813 Santa Barbara Street, 
  Pasadena, CA 91101, USA}

\altaffiltext{7}{Department of Astrophysics, Oxford University, Keble
  Road, Oxford, OX1 3RH, UK}

\begin{abstract}
We investigate the stellar populations of a sample of 162 Ly$\alpha$
emitting galaxies (LAEs) at $z = 3.1$ in the Extended Chandra Deep
Field South, using deep Spitzer IRAC data available from the GOODS and
SIMPLE surveys to derive reliable stellar population estimates.  We
divide the LAEs according to their rest-frame near-IR luminosities
into IRAC-detected and IRAC-undetected samples.  About 70\% of the
LAEs are undetected in 3.6 \micron\ down to $m_{3.6} = 25.2$ AB.
Stacking analysis reveals that the average stellar population of the
IRAC-undetected sample has an age of $\sim 200$ Myr and a mass of
$\sim \pow{3}{8}$ \Msun, consistent with the expectation that LAEs are
mostly young and low-mass galaxies.  On the other hand, the
IRAC-detected LAEs are on average significantly older and more
massive, with an average age $\ga 1$ Gyr and mass $\sim 10^{10}$
\Msun.  Comparing the IRAC colors and magnitudes of the LAEs to $z
\sim 3$ Lyman break galaxies (LBGs) shows that the IRAC-detected LAEs
lie at the faint blue end of the LBG color-magnitude distribution,
suggesting that IRAC-detected LAEs may be the low mass extension of
the LBG population.  We also present tentative evidence for a small
fraction ($\sim 5\%$) of obscured AGN within the LAE sample.  Our
results suggest that LAEs posses a wide range of ages and masses.
Additionally, the presence of evolved stellar populations inside LAEs
suggests that the Ly$\alpha$ luminous phase of galaxies may either be
a long-lasting or recurring phenomenon.
\end{abstract}


\section{Introduction} \label{Intro}
Lyman-alpha emitting galaxies (LAEs) are important tracers of galaxy
formation.  The Ly$\alpha$ emission is produced by on-going star
formation in the galaxies, and the line emission enables discovery of
objects that may be too faint to be seen in the continuum.  LAEs
therefore offer an opportunity to probe the faint end of galaxy
formation at high redshift, and may serve as building blocks of larger
galaxies in a hierarchical universe.

In recent years, there have been great advancements in narrow-band
surveys of LAEs.  Large samples of candidate LAEs now exist at
redshifts $z = 3 - 7$ \citep[e.g.][]{hu04, malho04, tanig05, shima06,
dawso07, muray07, ouchi07}.  One outstanding question regarding the
LAEs is their stellar population, which is important for understanding
the physical nature of the LAEs and their connection with other high
redshift galaxy populations such as Lyman break galaxies
\citep[LBGs;][]{shapl03, ando06, pente07, stanw07}.  A difficulty
often encountered when studying the stellar population of
high-redshift galaxies is the necessity to have observations at or
redward of rest-frame optical in order to constrain the total stellar
mass.  At $z \ga 3$, this means having either deep near-IR or Spitzer
IRAC observations.  Much progress has been made recently, however, and
several studies have suggested that LAEs are small galaxies with
masses $\la 10^9 - 10^{10}$ \Msun\ \citep{gawis06b, lai07, finke07,
pirzk07, nilss07}.

In this study, we focus on a sample of 162 LAEs at $z = 3.1$
discovered in a narrow-band survey of the Extended Chandra Deep Field
South (ECDF-S) by \citet{gronw07}.  The narrow-band 4990 \AA\ imaging
covers an area of 992 arcmin$^2$ and reaches a completeness limit of
$\sim \pow{1.5}{-17}$ erg cm$^{-2}$ s$^{-1}$ (equivalent to a
Ly$\alpha$ luminosity $L_{{\rm Ly}\alpha} = \pow{1.3}{42}$ erg
s$^{-1}$).  Spectroscopic results confirm the robustness of the sample
(details of the spectroscopy will be presented in \citealt{gawis07}
and Lira et al., in preparation).  Of the 52 candidates with
sufficient S/N in the spectra to yield redshifts, 51 were confirmed to
be $z=3.1$ LAEs (the remaining object is a $z=1.60$ AGN with C~III]
1909 \AA\ emission).  The most likely contaminants in our sample are
$z=0.34$ [O~II] emitters.  However, there should be a negligible
number of these objects in our sample because by requiring the
observer's frame equivalent width to be $> 80$ \AA, we eliminate all
but the strongest (and rarest) [O~II] emitters \citep{gronw07}.

The narrow-band imaging in ECDF-S is supplemented by optical and
near-IR observations from MUSYC \citep{gawis06a}, GOODS
\citep{dicki03}, and GEMS \citep{rix04} surveys, and Spitzer
observations from GOODS and SIMPLE\footnotemark\ (Damen et al., in
preparation).  Using the multi-wavelength data available in ECDF-S, we
study the rest-frame UV to near-IR properties of this large sample of
LAEs.  We place special emphasis on the new Spitzer IRAC observations,
which sample the rest-frame $0.9 - 2$ \micron\ emission from the $z =
3.1$ LAEs and provide valuable constraints on their stellar population
and star formation history.
\footnotetext{http://www.astro.yale.edu/dokkum/SIMPLE/}

We assume a ${\rm \Lambda}$CDM cosmology with $\Omega_{\rm M} = 0.3$,
$\Omega_{\rm \Lambda} = 0.7$, and $h = 0.7$.  All magnitudes are in
the AB system.

\section{IRAC Observations of $\zee = 3.1$ LAEs} \label{Analysis}
When working with deep IRAC observations, source confusion is often an
issue because of the broad IRAC PSF ($\sim 2\arcsec$ FWHM).  Since the
LAEs are in general quite faint in the IRAC bands, it is important to
select a subsample whose IRAC photometry is not significantly
contaminated by nearby objects.  Beginning with the original sample of
162 LAEs, we select those with no detected neighbors within a
4\arcsec\ radius ($\sim 2\times$ the IRAC FWHM), and additionally
reject LAEs with very bright neighbors at $> 4\arcsec$ radius by
visual inspection.  This results in a high confidence
IRAC-uncontaminated sample of 76 LAEs.

\begin{figure}
\includegraphics[width=\columnwidth]{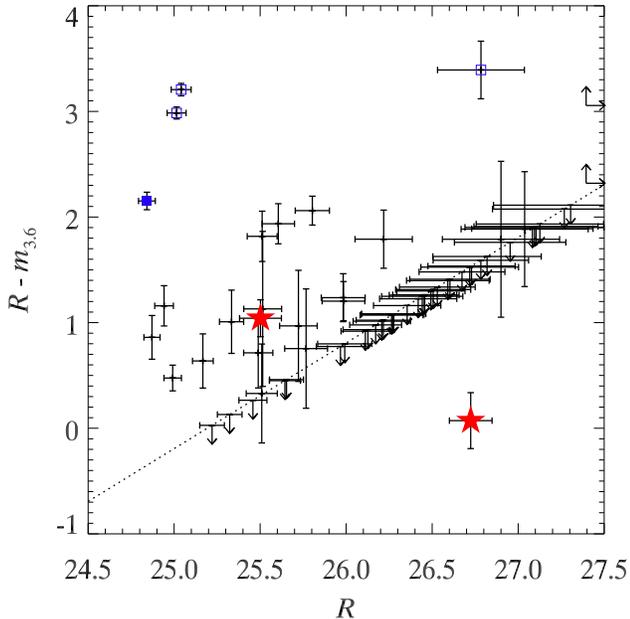}
\caption{ $R - m_{3.6}$ vs.\ $R$ color-magnitude diagram of the
  IRAC-uncontaminated LAE sample.  The dotted line represents the
  $m_{3.6}=25.2$ flux cut used to define the IRAC-undetected sample
  (shown as upper limits).  Objects that are undetected in both $R$
  and 3.6 \micron\ (15 in total) are omitted.  Note also that there
  are two objects that are detected in 3.6 \micron\ but not in $R$
  (shown as lower limits near the right of the plot).  Red $R -
  m_{3.6}>2$ objects are marked by blue squares, with the x-ray
  detected AGN candidate shown with a filled square.  The red
  five-point stars are results of stacking (in order of increasing
  $R$) the IRAC-detected and IRAC-undetected samples.
  \label{Rm36}
}
\end{figure}

We further separate the IRAC-uncontaminated sample into an
IRAC-detected sample and a corresponding IRAC-undetected sample.  This
division separates the rest-frame near-IR luminous objects from the
less luminous ones.  Of the 76 IRAC-uncontaminated LAEs, 52 are
undetected in the IRAC 3.6 \micron\ channel down to a flux limit of
0.3 $\mu$Jy ($m_{3.6}=25.2$).  This is nominally the 2$\sigma$
detection limit of the survey.  However, since the LAE positions are
well determined from the narrow band imaging, it is possible to do
slightly better with aperture photometry.  Aperture corrected fluxes
are measured from the IRAC images using 2\farcs5 diameter apertures.
The flux cut of 0.3 $\mu$Jy corresponds approximately to a 3$\sigma$
detection limit for the individual objects.  The initial IRAC-detected
sample consists of 24 objects, but 6 objects are rejected as possible
AGN or dusty objects (see below), leaving a total of 18 objects in the
IRAC-detected sample.

IRAC samples rest-frame $0.9-2$ \micron\ at $z \sim 3$ and provides
important constraints on the mature stellar population in the LAEs.
In particular, the $R-m_{3.6}$ color, shown for the
IRAC-uncontaminated sample in \Fig{Rm36}, probes the relative
contribution from young and mature stellar populations, and is
therefore an age indicator.  The dotted line represents the
$m_{3.6}=25.2$ flux limit for the IRAC-undetected sample.  All objects
in this sample will fall as upper limits on the dotted line.  The
survey limits imply that we are not sensitive to young stellar
populations with $R-m_{3.6} < 0$.  However, stacking analysis can be
used to obtain constraints on the average properties of the
IRAC-undetected sample (see next section).  The IRAC-detected LAEs
typically have red rest-frame UV to near-IR colors, $0 < R-m_{3.6} <
2$, with a median of $R-m_{3.6}=1.2$.  The red $R-m_{3.6}$ colors of
the IRAC-detected LAEs imply the possible presence of a significant
mature stellar population in LAEs, a point we will examine more
closely in the following section.

\begin{figure}
\includegraphics[width=\columnwidth]{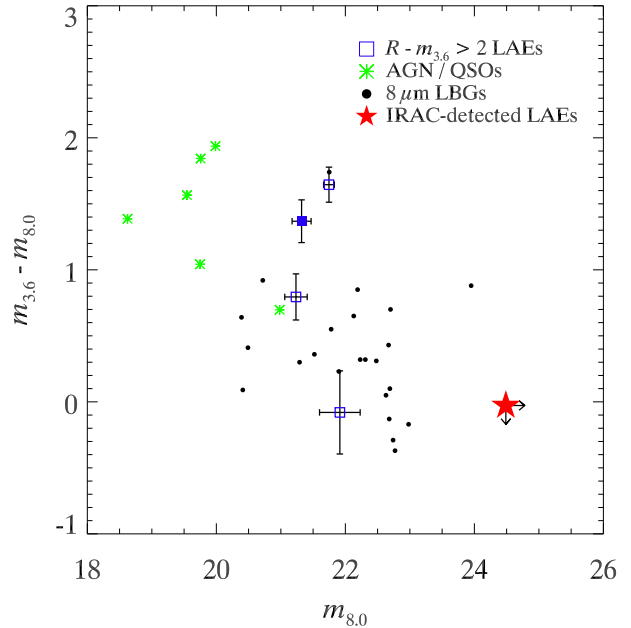}
\caption{ $m_{3.6}-m_{8.0}$ vs.\ $m_{8.0}$ color-magnitude diagram of
  the $R-m_{3.6} > 2$ LAEs (blue squares; the x-ray detected object is
  marked with a filled square), $z \sim 3$ AGN/QSOs (green stars;
  \citealt{steid03}; Magdis et al., in preparation), and 8 \micron\
  detected LBGs (black dots; \citealt{rigop06}).  The limits obtained
  by stacking the IRAC-detected sample are marked with the red star.
  \label{AGN_color}
}
\end{figure}

While the $R-m_{3.6}$ color traces the age of the stellar population,
the 3.6 \micron\ magnitude probes the mature stellar mass in the LAEs.
Therefore, the IRAC-detected sample, which has 3.6 \micron\ magnitudes
ranging from $m_{3.6} \sim 23.5 - 25$, should in general be more
massive than the $m_{3.6} > 25.2$ IRAC-undetected sample.  The
IRAC-detected sample is also in general brighter in $R$, with the
majority of IRAC-detected LAEs being brighter than $R \sim 26$.

We also explore the possibility of a correlation between Ly$\alpha$
luminosity and IRAC flux.  The mean Ly$\alpha$ luminosity of the
IRAC-detected sample ($\log(L_{{\rm Ly}\alpha}) = 42.47$ erg s$^{-1}$)
is about 40\% higher than that of the IRAC-undetected sample
($\log(L_{{\rm Ly}\alpha}) = 42.34$ erg s$^{-1}$).  This is only a
weak correlation, considering that the average 3.6 \micron\ flux of
the IRAC-detected sample is $8\times$ higher than the IRAC-undetected
sample (\Fig{sed}).  There is also significant overlap in the $L_{{\rm
Ly}\alpha}$ distributions of the two samples, in part due to the large
($\sim 0.2$ dex) dispersion in $L_{{\rm Ly}\alpha}$ within the
subsamples.

It is interesting to note that there are four objects in the sample,
corresponding to LAE 1, 11, 117, and 144 in the \citet{gronw07}
catalog, with very red colors $R-m_{3.6} > 2$.  LAE 1, the brightest
Ly$\alpha$ source in the sample, is detected in x-ray and is
presumably an AGN \citep{gronw07}.  The similar $R-m_{3.6}$ colors of
the remaining three x-ray undetected objects suggest that they may
possibly be AGN as well.  LAE 1, 11, and 117 are spectroscopically
confirmed to be at $z \sim 3$.  The x-ray detected object (LAE 1)
shows C~IV emission in its spectrum, while LAE 11 and 117 show only
Ly$\alpha$ emission (Lira et al., in preparation).  LAE 144 is
observed, but no line was detected in its spectrum.  LAE 11 and 117
are in the central part of CDF-S where the $0.5-8$ keV (rest-frame
$0.1-2$ keV) flux limit of $< \pow{3}{-16}$ erg s$^{-1}$ cm$^{-2}$
implies an upper limit of $<\pow{2.5}{43}$ erg s$^{-1}$ for the
objects' AGN x-ray luminosities.  The four objects have $R-m_{4.5}$
colors ranging from $2.4<R-m_{4.5}<4$, consistent with the colors of
mid-IR selected AGN \citep{hicko07}.  However, the observed
$R-m_{4.5}$ colors can also be produced by a dusty stellar continuum
with $E(B-V) \sim 0.5$.  These four objects are the only 8.0 \micron\
detected sources in the IRAC-uncontaminated sample and with $m_{8.0}$
ranging from $21 - 22$ they are more than 2 mags brighter than the
average IRAC-detected LAE (\Fig{AGN_color}).  Two of the four objects
have $m_{3.6} - m_{8.0}$ colors consistent with $z \sim 3$ AGN and
QSOs (\citealt{steid03}; Magdis et al., in preparation).  However,
\Fig{AGN_color} shows that the four objects have 8.0 \micron\
magnitudes more similar to that of the 8 \micron\ detected LBGs
\citep{huang05, rigop06}.  There are a total of two x-ray detected
objects in the original sample of 162 LAEs (the other x-ray detected
object is not in the IRAC-uncontaminated sample), giving an AGN
fraction of $\sim 1\%$.  The evidence presented here cannot rule out
the possibility that the other three red $R-m_{3.6}$ LAEs are dusty
LBG-like objects, but if they also turn out to be AGN, then the AGN
fraction is $\sim 5\%$.

Since the four $R-m_{3.6} > 2$ LAEs have rather different properties
than the rest of the LAE population, they are excluded from the
conservative IRAC-detected sample.  Two more objects are rejected
since they are detected in 3.6 \micron\ but not in $R$, which may
imply a large amount of dust reddening inconsistent with the rest of
the IRAC-detected population.  The subsequent analysis will be focused
on the conservative IRAC-detected sample of 18 LAEs.  However, we will
also present results derived using the larger sample of 23
IRAC-detected LAEs (the x-ray detected LAE AGN is still excluded).

\section{Stellar Population} \label{SP}
Since the majority of the LAEs in our sample are undetected in the
IRAC images, we perform a stacking analysis in order to measure the
average properties of the sample.  Again, we work only with the sample
of 76 IRAC-uncontaminated LAEs.  This is especially important when
stacking the IRAC-undetected subsample because any flux contributions
from bright neighbors could significantly skew the average.  Because
of crowdedness in the IRAC images, it is also necessary to clean the
images before stacking by subtracting detected sources using a PSF
fitting routine \citep[StarFinder;][]{diola00}.  The optical and
near-IR images are stacked in the same manner as the IRAC images,
except that the subtraction of detected sources is not necessary in
this case.  Uncertainties in the stacked photometry are estimated by
bootstrapping the original sample, and then repeating the stacking and
flux measurement on the bootstrap samples.  The uncertainties thus
derived take into account both random fluctuations and the intrinsic
luminosity dispersion among the LAE population.

\begin{figure}
\includegraphics[width=\columnwidth]{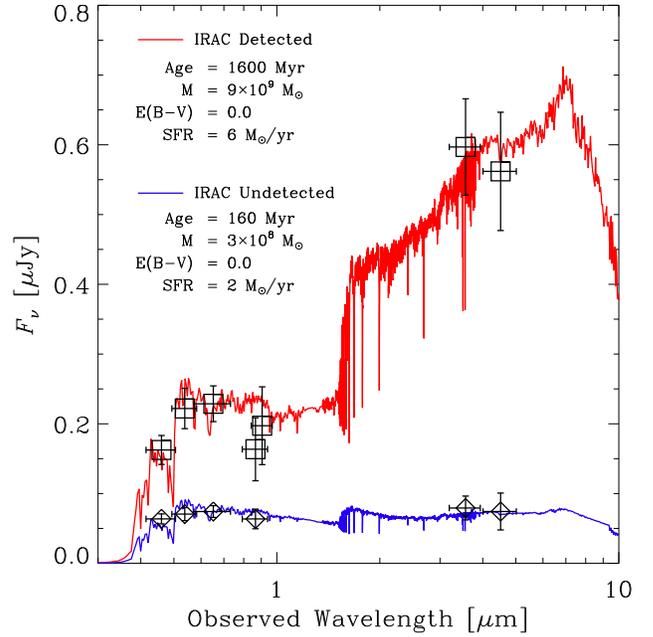}
\caption{ Stacked SEDs and best-fit models of the IRAC-detected
  (squares, red line) and IRAC-undetected (diamonds, blue line)
  samples.  The models assume constant SFR and solar metallicity.
  Only the bands with detections in the stacked images (${\rm S/N} >
  3$) are included in the plot.
  \label{sed}
}
\end{figure}

We fit \citet{bc03} stellar population synthesis models with a
Salpeter initial mass function to the stacked photometry.  The $V$
band flux is corrected for the contribution of the Ly$\alpha$ emission
before the fits.  The best-fit model to the IRAC-undetected sample is
shown together with the stacked SED in \Fig{sed}.  Assuming a constant
SFR and solar metallicity, the IRAC-undetected sample is best fitted
by a model with $E(B-V) = \err{0.0}{0.0}{0.1}$, age =
\err{160}{110}{140} Myr, and mass = $\err{3}{2}{4}\times 10^8$ \Msun.

The average mass of the IRAC-undetected LAEs is smaller than that of
typical LBGs by more than a factor of 30.  The average SFR derived
from the best-fit model is 2 \Msun\ yr$^{-1}$, which is not very large
when compared to other populations of galaxies at $z \sim 3$ such as
the distant red galaxies (DRGs; \citealt{franx03}) which typically
have SFR $\ga 100$ \Msun\ yr$^{-1}$ \citep{forst04}.  However, the
LAEs have a specific SFR (defined as SFR per unit mass) of \pow{7}{-9}
yr$^{-1}$, which is among the highest observed of galaxies at $z \sim
3$.  In comparison, LBGs have specific SFR $\sim \pow{3}{-9}$
yr$^{-1}$ \citep{shapl01} and DRGs have specific SFR $\sim
\pow{1}{-9}$ yr$^{-1}$ \citep{forst04}.  This means that the stellar
populations in LAEs are increasing in mass at a much faster rate than
other populations of galaxies at similar redshifts, suggesting that
they are in the early stages of formation.

The IRAC-detected sample appears to be significantly different from
the IRAC-undetected sample (\Fig{sed}).  Again assuming a constant SFR
and solar metallicity, we find that the average stellar population of
the IRAC-detected LAEs is best fitted by a model with $E(B-V) =
\err{0.0}{0.0}{0.1}$, age = $1.6 \pm 0.4$ Gyr, and mass = $9 \pm 3
\times 10^9$ \Msun.  Compared to the IRAC-undetected sample, the SFR
of the IRAC-detected LAEs is slightly higher at 6 \Msun\ yr$^{-1}$.
The specific SFR, however, is about $10 \times$ lower at \pow{7}{-10}
yr$^{-1}$, similar to LBGs with comparable stellar population ages
\citep{shapl01}.  We emphasize that although the IRAC-detected and
IRAC-undetected samples have distinct properties on average, the
individual LAEs exhibit a continuum of properties, as evident in
\Fig{Rm36} \citep[see also][]{gronw07}.  If we include the 5 red
$R-m_{3.6}$ objects originally excluded from our conservative
IRAC-detected sample (the x-ray detected LAE AGN is still excluded in
this analysis), the best-fit age and mass increase slightly to 2.1 Gyr
and \pow{1.8}{10} \Msun, respectively.

The best-fit parameters presented above do not depend sensitively on
the assumed metallicity.  Fitting the data with a 0.2 \Zsun\ model
increases the best-fit mass and the age of the IRAC-undetected sample
by 30\% and 60\%, respectively, while the effect on the IRAC-detected
sample is even smaller at less than 10\%.  The best-fit parameters
also do not depend sensitively on the assumed star formation history.
Comparing constant and exponentially declining SFR models with the
same metallicity, we find that the two models in general give best-fit
masses and ages that are within $\sim 20\%$ of each other.  This is
because the best-fit exponential SFR models have mostly long star
formation timescales ($\ga 1$ Gyr) and are therefore very similar to a
constant SFR model.

At $z = 3.1$, AGB stars may contribute significantly to the flux in
the IRAC bands, which correspond to rest-frame 0.9 - 2 \micron.  In
order to examine how the effects of AGB stars would change our
conclusions, we perform alternate fits using \citet{maras05} stellar
population synthesis models, which include larger contributions from
AGB stars than \citet{bc03} models.  Fitting exponentially declining
SFR models with solar metallicity, we find for the IRAC-undetected
sample a best-fit model with age = 100 Myr, star formation timescale =
250 Myr, $E(B-V) = 0$, and mass = \pow{3}{8} \Msun.  For the
IRAC-detected sample, the best-fit model has age = 700 Myr, star
formation timescale = 2 Gyr, $E(B-V) = 0$, and mass = \pow{5}{9}
\Msun.  The Maraston models in general give lower mass and age
estimates, but despite these differences, the distinction between the
IRAC-detected and IRAC-undetected populations remains significant.

\section{Discussion} \label{Disc}
There are several previous studies on the stellar population of LAEs
at $z \sim 3$ and beyond \citep{gawis06b, finke07, pirzk07, nilss07}.
These previous studies found that LAEs have masses ranging from $10^6
- 10^9$ \Msun, and ages ranging from $0.005 - 1$ Gyr.  Our best-fit
average age of 160 Myr and mass of \pow{3}{8} \Msun\ for the
IRAC-undetected sample are broadly consistent with these previous
studies.  In a further analysis of the IRAC-undetected sample
presented in this paper, \citet[submitted]{gawis07} fit two-component
models and find that the IRAC-undetected LAEs have a total mass of
\pow{1}{9} \Msun, with a young stellar component (age $\sim 20$ Myr)
accounting for $\sim 20\%$ of the total mass.  These results are
consistent with ours which are derived using single-component models
and should be interpreted as the average properties of the entire
stellar population within the LAEs.

\begin{figure}
\includegraphics[width=\columnwidth]{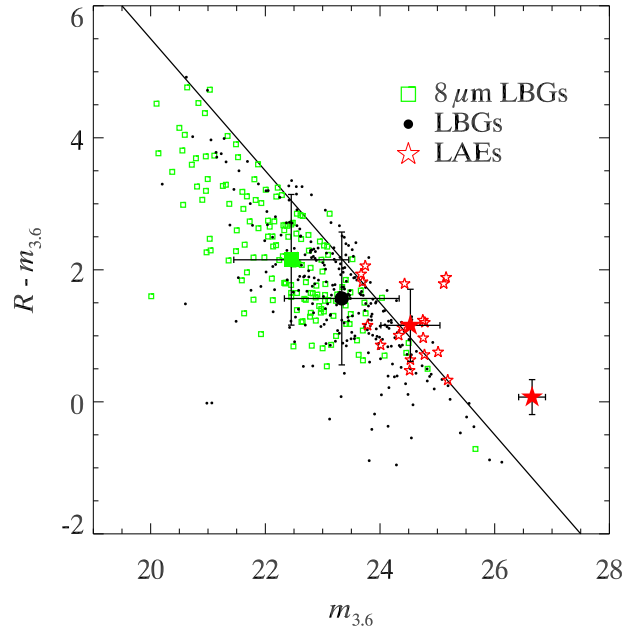}
\caption{ A comparison of the $R-m_{3.6}$ colors and 3.6 \micron\
  magnitudes between 8 \micron\ detected LBGs (green square), regular
  LBGs (black dots), and the IRAC-detected LAEs (red stars).  The
  solid line shows the $R<25.5$ selection limit for LBGs.  The data
  for the regular LBGs come from IRAC observations of the
  \citet{steid03} LBG sample (Magdis et al., in preparation), and the
  8 \micron\ LBG data come from combining the samples of Magdis et
  al.\ and \citet{rigop06}.  The mean colors and magnitudes of each
  sample are plotted as large filled symbols, with error bars showing
  the dispersion within the samples.  We additionally plot the stacked
  photometry of the IRAC-undetected sample, shown as the red filled
  red star with the faintest 3.6 \micron\ magnitude.
  \label{compLBG}
}
\end{figure}

The IRAC-detected LAEs are more luminous than the IRAC-undetected LAEs
in both rest-frame UV and near-IR, and represent the massive end of
the LAE mass spectrum.  Potentially, the IRAC-detected LAEs may
provide a link between the LAEs and other galaxy populations.  For
instance, they may be the $z \sim 3$ analogues to the similarly
massive IRAC-detected LAEs found at $z \sim 5.7$ \citep{lai07}.  In
\Fig{compLBG}, we compare the $R-m_{3.6}$ colors and 3.6 \micron\
magnitudes of the LAEs to 8 \micron\ detected LBGs and regular LBGs at
$z \sim 3$.  The $R-m_{3.6}$ color and 3.6 \micron\ magnitude
correlate with stellar mass \citep{rigop06}, with the 8 \micron\ LBGs
being the most massive ($\ga 10^{11}$ \Msun), followed by LBGs ($\sim
10^{10} - 10^{11}$ \Msun), IRAC-detected LAEs ($\sim 10^{10}$ \Msun),
and finally IRAC-undetected LAEs ($\sim 10^8$ \Msun).  The
IRAC-detected LAEs occupy the faint and blue end of the LBG
color-magnitude distribution, suggesting that they may be the lower
mass extension of the LBG population.  This interpretation is
supported by the fact that the inferred mass of the IRAC-detected LAEs
($\sim 10^{10}$ \Msun) is at the low end of the mass range found for
LBGs ($\sim 10^{10} - 10^{11}$ \Msun; \citealt{papov01, shapl01}).
Furthermore, most LAEs in the present sample have (observer frame)
optical colors that would allow them to be selected as LBGs, although
the LAEs tend to be fainter in the rest-frame UV \citep{gronw07,
gawis06b}.  It has also been observed that $\sim 20\% - 25\%$ of LBGs
at $z \sim 3$ exhibit Ly$\alpha$ emission strong enough to be narrow
band excess objects \citep{steid00, shapl03}.  There is thus
accumulating evidence suggesting that the IRAC-detected LAEs may be
the bridge connecting the LAE and LBG populations.

Using the stellar mass estimates derived in the previous section and
assuming a survey volume of \pow{1.1}{5} Mpc$^3$ \citep{gronw07}, we
find stellar mass densities of $0.3 \pm 0.3 \times 10^6$ and $3 \pm 1
\times 10^6$ \Msun\ Mpc$^{-3}$ for the IRAC-undetected and
IRAC-detected populations, respectively.  The errors in the stellar
mass densities include uncertainties in the mass from the fits and
uncertainties in the number density coming from Poisson fluctuations
and a $\sim 20\%$ cosmic variance given the observed LAE bias of $\sim
1.7$ \citep[submitted]{gawis07}.  We should stress that the above
stellar mass densities only account for LAEs with Ly$\alpha$
luminosities above our survey completeness limit of \pow{1.3}{42} erg
s$^{-1}$.  The IRAC-undetected LAEs account for about $9 \pm 10\%$ of
the total stellar mass in LAEs, even though they make up 2/3 of the
total by number.  Compared to LBGs and DRGs at $z \sim 3$, which have
stellar mass densities of $\sim \pow{1}{7}$ and \pow{7}{6} \Msun\
Mpc$^{-3}$ respectively \citep{grazi07}, the stellar mass contained in
LAEs is smaller, but not insignificant.  However, it is important to
keep in mind that there may be substantial overlap between the LAEs
and LBGs.  A better understanding of the overlap between these two
populations is necessary before a direct comparison of the stellar
mass densities can be made.

The present results show that the LAEs posses a wide range of masses
and ages, from the massive and evolved IRAC-detected LAEs to the young
and small IRAC-undetected LAEs.  The range of photometric properties
shown in the LAE sample suggests that LAEs exhibit a continuum of
properties between these two extremes.  The presence of both young and
evolved stellar populations within the overall LAE population implies
that the Ly$\alpha$ luminous phase of galaxies may last $\ga 1$ Gyr,
or that the Ly$\alpha$ luminous phase is recurring.  Interestingly,
\citet{shapl01} found that LBGs with best-fit stellar population ages
$\ga 1$ Gyr also show strong Ly$\alpha$ emission in their spectra.
The authors suggest that vigorous past star formation has destroyed
and/or expelled the dust inside the galaxies, allowing the Ly$\alpha$
photons to escape.  One characteristic of these evolved Ly$\alpha$
emitting LBGs is that they tend to have more quiescent SFR than the
rest of the population.  The IRAC-detected LAEs, with their evolved
stellar population and low specific SFR, are consistent with this
scenario.  If the LAE phase is recurring, then the young age of the
IRAC-undetected LAEs implies that their number and stellar mass
densities can be as much as a factor $\sim 10$ higher (very roughly
the ratio of the age of the universe to the stellar age).  Similarly
young stellar populations ($\la 100$ Myr) have also been found at
higher redshifts \citep{finke07, pirzk07, verma07}.  The stochastic
nature of galaxies with short lifetimes suggests that there may be a
related population of undetected pre or post-starburst galaxies that
may contribute significantly to the stellar mass and star formation
rate densities at $z \ga 3$.

\acknowledgements

We thank the referee for comments that helped improve the paper.  KL
thanks Ryan Hickox for helpful discussions.  This work is based in
part on archival data obtained with the Spitzer Space Telescope, which
is operated by the Jet Propulsion Laboratory, California Institute of
Technology under a contract with NASA.  Support for this work was
provided by NASA through an award issued by JPL/Caltech.

\bibliographystyle{apj}
\bibliography{ms}

\end{document}